\documentclass[aps,pre,reprint,showpacs]{revtex4-1}

\usepackage{amsmath,amssymb,amsthm}
\usepackage[dvipdfmx]{graphicx,hyperref}
\usepackage{braket}

\newcommand{\tr}{\mathrm{tr}}

\begin{document}

\title{Work Extraction from a Single Energy Eigenstate}

\author{Kazuya Kaneko}
\affiliation{
Department of Applied Physics, The University of Tokyo,
7-3-1 Hongo, Bunkyo-ku, Tokyo 113-8656, Japan
}

\author{Eiki Iyoda}
\affiliation{
Department of Applied Physics, The University of Tokyo,
7-3-1 Hongo, Bunkyo-ku, Tokyo 113-8656, Japan
}

\author{Takahiro Sagawa}
\affiliation{
Department of Applied Physics, The University of Tokyo,
7-3-1 Hongo, Bunkyo-ku, Tokyo 113-8656, Japan
}

\date{\today}

\begin{abstract}
Work extraction from the Gibbs ensemble by a cyclic operation
is impossible, as represented by the second law of thermodynamics.
On the other hand,
the eigenstate thermalization hypothesis (ETH) states that
just a single energy eigenstate can describe a thermal equilibrium state.
Here we attempt to unify these two perspectives
and  investigate the second law at the level of individual energy eigenstates,
by examining the possibility of extracting work from a single energy eigenstate.
Specifically,
we performed numerical exact diagonalization of a quench protocol of local Hamiltonians
and evaluated the number of work-extractable energy eigenstates.
We found that it becomes exactly zero in a finite system size,
implying that a positive amount of work cannot be extracted from any energy eigenstate,
if one or both of the pre- and the post-quench Hamiltonians are non-integrable. 
We argue that the mechanism behind this numerical observation
is based on the ETH for a non-local observable.
Our result implies that quantum chaos,
characterized by non-integrability,
leads to a stronger version of the second law
than the conventional formulation based on the statistical ensembles.
\end{abstract}


\maketitle

\section{Introduction}
How much work can we extract from an isolated many-body quantum system?
This question brings us to the foundation of the second law of thermodynamics
\cite{Tasaki2000stat,Allahverdyan2002}.
In fact,
a standard way of representing the second law is that
we cannot extract a positive amount of work by any cyclic process,
which is nothing but the impossibility of the perpetual motion of the second kind.
Here, \textit{cyclic} means that the initial and final Hamiltonians are the same.
From the viewpoint of quantum statistical mechanics,
a quantum state is called passive,
if a positive amount of work cannot be extracted from that state by any cyclic process
\cite{Pusz1978,Lenard1978}.
For example,
the Gibbs states are passive,
while pure states, except for the ground states,
are not passive.

In a slightly different context,
thermalization in isolated quantum systems
has recently attracted renewed interest
\cite{Polkovnikov2011,DAlessio2015,Eisert2015,Gogolin2016}.
Recent studies have revealed that
a thermal equilibrium state can be represented
not only by a statistical ensemble
such as the Gibbs state,
but also by a single energy eigenstate;
this is called the eigenstate thermalization hypothesis (ETH)
\cite{Berry1977,Peres1984,Deutsch1991,Srednicki1994,Rigol2008}.
It has been established
that the non-integrability of the Hamiltonian
is crucial for the strong version of the ETH,
which states that \textit{every} energy eigenstate is thermal
\cite{Jensen1985,Rigol2008,Steinigeweg2013,Steinigeweg2014,Beugeling2014,Kim2014,Khodja2015,Garrison2015,Mondaini2016,Dymarsky2018,Yoshizawa2017}.
On the other hand, for integrable systems,
the strong ETH fails
\cite{Jensen1985,Rigol2008,Rigol2009,Biroli2010,Steinigeweg2013,Alba2015,Essler2016,Vidmar2016,Yoshizawa2017},
but only a weaker version of the ETH holds,
which states that \textit{most} of energy eigenstates are thermal
\cite{Biroli2010,Alba2015,Tasaki2015,Iyoda2016,Mori2016,Yoshizawa2017}.

Motivated by these modern researches of thermalization,
the second law for quantum pure states
has been discussed recently
\cite{Tasaki2000pure,Goldstein2013,Ikeda2013,Iyoda2016,Kaneko2017}.
However,
the aforementioned original question,
i.e., the possibility of work extraction from an isolated many-body quantum system
\cite{Dorner2013,Gallego2014,Perarnau-Llobet2016,Modak2017,Le2018},
has not been fully addressed in light of the ETH
\cite{Jin2016,Schmidtke2017}.
Since a single energy eigenstate is not passive,
any amount of work can be extracted from it,
if any cyclic unitary operation is allowed
\cite{Pusz1978,Lenard1978}.
Therefore, a natural question is
whether work extraction is possible or not by \textit{physical} Hamiltonians with local interactions,
and whether that possibility depends on the integrability of the Hamiltonian or not.

In this paper,
we investigate work extraction from a single energy eigenstate. 
By using numerical exact diagonalization,
we found that
for a quench protocol of local Hamiltonians,
one cannot extract a positive amount of work from \textit{any} energy eigenstate,
if one or both of the pre- and the post-quench Hamiltonians are non-integrable.
This is a strong version of the second law,
which can be referred to as the ``eigenstate'' second law
and is analogous to the strong ETH.
In fact,
we argue that this version of the second law follows
from the strong ETH of a non-local many-body observable,
which itself is nontrivial.
On the other hand,
if both the pre- and the post-quench Hamiltonians are integrable,
this stronger version of the second law fails.
To complement our numerical observations, 
we analytically derive a weaker version of  the eigenstate second law,
stating that for any unitary operation, 
one cannot extract a positive amount of work from \textit{most} of energy eigenstates.
This argument is true regardless of the integrability of the Hamiltonian,
and is analogous to the weak ETH.

Our results suggest that
the second law of thermodynamics is true
even at the level of individual energy eigenstates
if the system is non-integrable
(i.e., quantum chaotic).
Our work would serve as a foundation for quantum many-body heat engines,
which can be experimentally realized by modern quantum technologies.
In fact, our setup
would be relevant to a variety of experimental systems,
including ultracold atoms and superconducting qubits
\cite{Trotzky2011,Langen2015,Neill2016}.

This paper is organized as follows.
In Sec.~\ref{sec:numerical},
we show our main numerical results.
In Sec.~\ref{sec:ETH},
we discuss the relationship between work extraction and
the strong ETH of a non-local observable.
In Sec.~\ref{sec:bound},
we analytically show a weaker version of  the eigenstate second law.
In Sec.~\ref{sec:conclusion},
we summarize our results.
In Appendix,
we provide supplemental numerical results.

\section{Main results}
\label{sec:numerical}
In this section, we discuss our setup and the main numerical results.

\subsection{Formulation}
\label{sec:formulation}
We first formulate work extraction from a quantum many-body system
by a cyclic operation
\cite{Pusz1978,Lenard1978}.
We drive the system by the time-dependent Hamiltonian $ H(t) $
from $ t=0 $ to $ t=\tau $.
The corresponding time evolution is represented by
the unitary operator
$ U=\mathcal{T}\exp\left(
-i\int_0^\tau H(t)dt
\right) $,
where $ \mathcal{T} $ is the time-ordering operator.
We assume that the initial and the final Hamiltonians are the same:
$ H(0)=H(\tau) =:H_{\mathrm{I}} $.

The average work extraction from the state $ \rho $ during this protocol is
\begin{align}
	W := \tr[H_{\mathrm{I}} \rho]-\tr[H_{\mathrm{I}} U\rho U^\dag].
\end{align}
We focus on work extraction from a single energy eigenstate of $  H_{\mathrm{I}}$.
Since an energy eigenstate is not passive,
$ W $ can be positive for some unitary operations.
Given that, in the following, we focus on
whether we can extract positive work by
quench protocols of local Hamiltonians,
and numerically investigate
the fundamental difference between integrable and non-integrable systems.

\subsection{Model and protocols}

As a simple model that we can control integrability,
we study the quantum Ising model
with a transverse field $ g$,
longitudinal field $ h $,
nearest-neighbor coupling $ J $,
and next nearest-neighbor coupling $ J' $.
The Hamiltonian is given by
\begin{align}
H=&\sum_{i=1}^N(g\sigma_i^x+h \sigma_i^z+J\sigma_i^z\sigma_{i+1}^z+J'\sigma_i^z\sigma_{i+2}^z),
\end{align}
where $ \sigma_i^x, \sigma_i^z  $
are the local Pauli operators on the site $ i $,
and $  N $ is the number of spins.
We adopted the periodic boundary condition.
This model is integrable when $ g=0 $ or $ h=J'=0 $.
In non-integrable cases, the level spacing statistics
can be well fitted by the Wigner-Dyson distribution
(see the Appendix \ref{sec:level_statistics} for $ N=18, J=1,J'=0, g=\frac{\sqrt{5}+5}{8}, 2.5\times 10^{-2}\lesssim h\lesssim 1.5 $).
This model is also known to satisfy
the strong ETH in the non-integrable parameter region
\cite{Kim2014,Dymarsky2018}.

We next describe the quench protocol in our numerical simulation.
Let $ H_{\mathrm{I}}  $ be the initial Hamiltonian,
and suppose that 
the system is initially prepared in its eigenstate $ \ket{j}$
with energy $ E_j $.
We perform the following work extraction protocol:
(i) At initial time $ t= 0 $,
the longitudinal field and  transverse field
are suddenly quenched
($ h_{\mathrm{I}}\to h_{\mathrm{Q}}, g_{\mathrm{I}}\to g_{\mathrm{Q}}$).
The post-quench Hamiltonian is denoted by $ H_{\mathrm{Q}} $.
(ii) The system evolves from time $0$ to $\tau$
according to the new Hamiltonian $ H_{\mathrm{Q}} $.
(iii) At time $t=\tau$,
we again quench the fields in the backward direction
($ h_{\mathrm{Q}}\to h_{\mathrm{I}}, g_{\mathrm{Q}}\to g_{\mathrm{I}}$)
and restore the Hamiltonian to $ H_{\mathrm{I}}  $.

The average work extraction from eigenstate $ \ket{j} $
during this protocol is given by 
\begin{align}
W_j=
E_j-\braket{j|e^{iH_{\mathrm{Q}} \tau}H_{\mathrm{I}}e^{-iH_{\mathrm{Q}} \tau} |j}.
\end{align}
We then consider the following
four cases of the integrability:
(a) $ H_{\mathrm{I}} $ is non-integrable and $ H_{\mathrm{Q}}  $ is non-integrable;
(b) $ H_{\mathrm{I}} $ is integrable and $ H_{\mathrm{Q}}  $ is non-integrable;
(c) $ H_{\mathrm{I}} $ is non-integrable and $ H_{\mathrm{Q}}  $ is integrable;
(d) $ H_{\mathrm{I}} $ is integrable and $ H_{\mathrm{Q}}  $ is integrable.

We note the definition of the inverse temperature of an energy eigenstate.
The inverse  temperature $ \beta_j $ associated
with an energy eigenstate $ \ket{j} $
is defined through the equation
$ E_j= u_\mathrm{Gibbs}(\beta_j) $,
where $ u_\mathrm{Gibbs}(\beta_j)$ is the energy density
of the Gibbs state at inverse temperature $ \beta_j $.
We cannot extract positive work form the Gibbs state
with positive inverse temperature $ \beta> 0 $,
but can do from that with negative inverse temperature $ \beta<0 $.

\subsection{Distribution of extracted work}

\begin{figure}
	\includegraphics[width=1.0\linewidth]{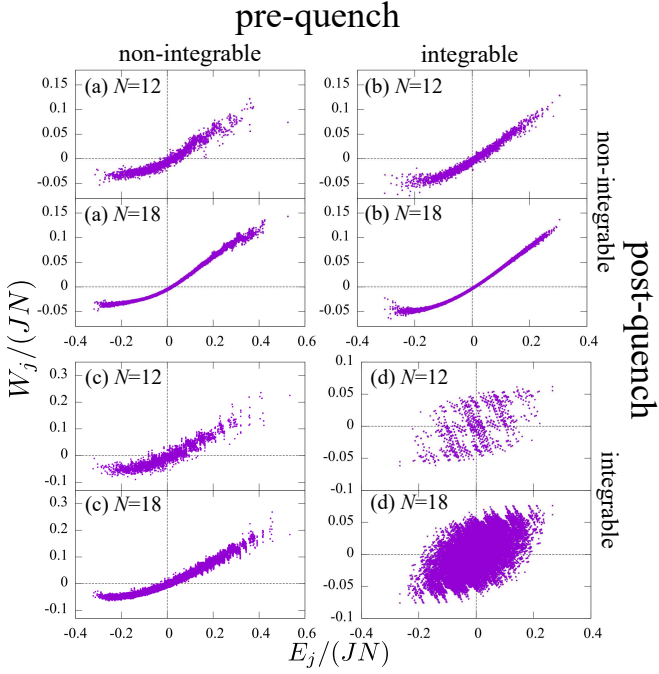}
	\caption{\label{fig:quench_Ising}
		The average work $W_j / N$ versus the eigenenergy $E_j / (JN)$
		in the full spectrum for $ N=12 $ (upper panels) 
		and $ N=18 $ (lower panels).
		The waiting time after the quench
		is $ J\tau=10^4 $ in all the cases.
		The vertical dashed line indicates the energy at $ \beta_j=0 $,
		and the horizontal one indicates $ W_j/(JN)=0 $.
		$\beta_j > 0$ in the left side of the vertical line, while $\beta_j < 0$ in the right side.
		Quench from 
		(a) non-integrable to non-integrable
		($ J=J'=1, g_{\mathrm{I}}=g_{\mathrm{Q}}=(\sqrt{5}+5)/8, h_{\mathrm{I}}=0\to h_{\mathrm{Q}}= 1.0 $),
		(b) integrable to non-integrable
		($ J=1, J'=0, g_{\mathrm{I}}=g_{\mathrm{Q}}=(\sqrt{5}+5)/8, h_{\mathrm{I}}=0\to h_{\mathrm{Q}}=1.0$),
		(c) non-integrable to integrable
		($ J=1, J'=0, g_{\mathrm{I}}=g_{\mathrm{Q}}=(\sqrt{5}+5)/8, h_{\mathrm{I}}=1.0 \to h_{\mathrm{Q}}=0 $),
		and
		(d) integrable to integrable
		($ J=1, J'=0, h=0, g_{\mathrm{I}}=0.5\to g_{\mathrm{Q}}=1.5 $).
	}
\end{figure}

We show our numerical result on
the distribution of the extracted work over energy eigenstates.
Figure \ref{fig:quench_Ising} shows
$W_j$ versus the eigenenergy $E_j$.
Overall,
the average work tends to be negative (positive) in the positive (negative) temperature region.
 
Specifically, in the case of (a), (b), and (c),
where one or both of Hamiltonians $ H_{\mathrm{I}} $ and $ H_\mathrm{Q} $ are non-integrable, 
the distribution of the data points becomes narrower in the horizontal direction
from $N=12$ to $18$.
We thus expect that
these data points converge to a single smooth curve
in the thermodynamic limit $N \to \infty$.
This is analogous to the strong ETH
\cite{Jensen1985,Rigol2008,Steinigeweg2013,Steinigeweg2014,Beugeling2014,Kim2014,Khodja2015,Garrison2015,Mondaini2016,Dymarsky2018,Yoshizawa2017}.
On the other hand,
in the case of (d),
the distribution of the data points is more spread out
than the other cases
and does not look like convergent to a single curve.
This is analogous to the behavior of the ETH in integrable systems
\cite{Rigol2008,Biroli2010}.

We also note that in the case of (a), (b), and (c),
eigenstates with $ W_j>0 $
do not exist in the positive temperature region
for $ N=18 $,
whereas
in the integrable case (d),
there are work extractable eigenstates also in the positive temperature region.

To confirm the validity of the above argument more quantitatively,
we systematically investigate 
the number of work-extractable energy eigenstates
in the next subsection.

\subsection{Dependence on size and integrability}

\begin{figure}
	\includegraphics[width=1.0\linewidth]{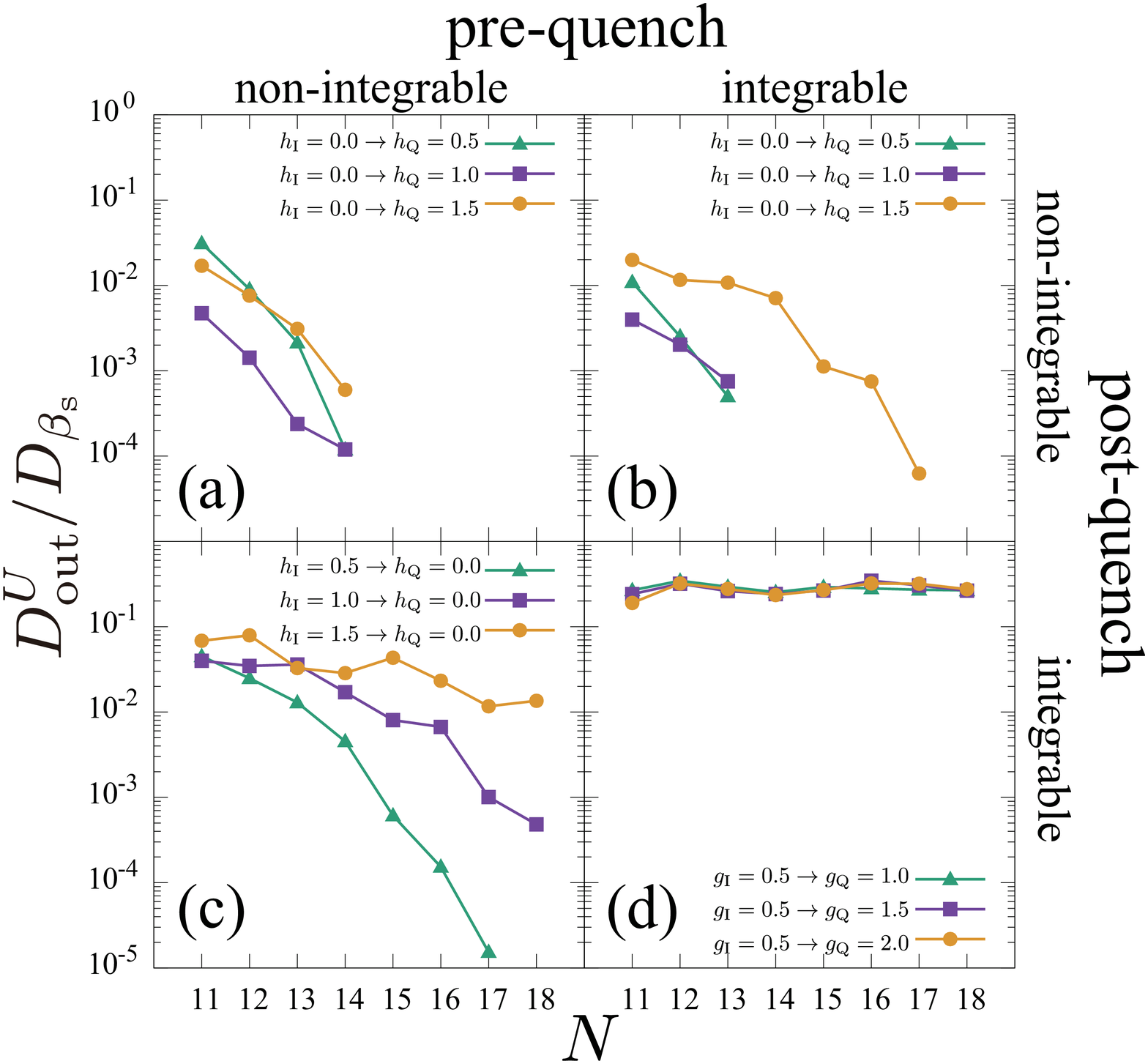}
	\caption{\label{fig:scaling}
		The system-size dependence of the ratio $ D^U_{\mathrm{out}}/D_{\beta_\mathrm{s}}    $.
		The lack of a data point means that
		$D^U_{\mathrm{out}}$ is exactly zero there.
		The waiting time after the quench is $ J\tau=10^4 $ in all the cases.
		We set the thresholds as $ J\beta_{\mathrm{s}}=0.02 $ and $ \delta/J=0.001 $.
		Quench from
		(a) non-integrable to non-integrable
		($ J=J'=1, g_{\mathrm{I}}=g_{\mathrm{Q}}=(\sqrt{5}+5)/8, h_{\mathrm{I}}=0\to h_{\mathrm{Q}}= 0.5, 1.0$, and $1.5 $),
		(b) integrable to non-integrable
		($ J=1, J'=0, g_{\mathrm{I}}=g_{\mathrm{Q}}=(\sqrt{5}+5)/8, h_{\mathrm{I}}=0\to h_{\mathrm{Q}}=0.5, 1.0$, and $ 1.5 $) ,
		(c) non-integrable to integrable
		($ J=1, J'=0, g_{\mathrm{I}}=g_{\mathrm{Q}}=(\sqrt{5}+5)/8, h_{\mathrm{I}}=0.5, 1.0, $ and $ 1.5 \to h_{\mathrm{Q}}=0 $),
		and
		(d) integrable to integrable
		($ J=1, J'=0, h=0, g_{\mathrm{I}}=0.5\to g_{\mathrm{Q}}=1.0, 1.5 $, and $ 2.0 $).
	}
\end{figure}

We next quantitatively study
the system-size dependence of the number of energy eigenstates
from which one can extract positive work.
We focus on eigenstates at positive finite temperature $ (\beta_j>0 ) $.
In addition,
we set a threshold for the inverse temperature,
denoted as $ \beta_\mathrm{s}>0 $,
because the finite-size effect is significant in small $ \beta_j $.
The set of energy eigenstates with $ \beta_j\geq\beta_\mathrm{s} $
is denoted by
$ 
M_{\beta_\mathrm{s}}
:=\{
j: \beta_j\geq \beta_\mathrm{s}
\}.
$
We denote the number of energy eigenstates in $ M_{\beta_\mathrm{s}} $
by $ D_{\beta_\mathrm{s} } $.
Let $ D^U_{\mathrm{out}} $
be the number of eigenstates in $ M_{\beta_\mathrm{s}}$
from which one can extract work greater than $ N\delta $
(i.e.,
$ D^U_{\mathrm{out}}
:=
\#\{
j\in M_{\beta_\mathrm{s}}
:
W_j>N\delta
\} $
).
We numerically investigate 
the system-size dependence of the ratio $ D^U_{\mathrm{out}}/D_{\beta_\mathrm{s}}  $.

Figure~\ref{fig:scaling}
shows the system-size dependence of $ D^U_{\mathrm{out}}/D_{\beta_\mathrm{s}}   $
for several quench protocols.
In the case of  (a), (b), and (c),
where one or both of Hamiltonians $ H_\mathrm{I} $ and $ H_\mathrm{Q} $ are non-integrable, 
$ D^U_{\mathrm{out}}/D_{\beta_\mathrm{s}}    $ decays at least exponentially with $N$.
On the other hand,
$ D^U_{\mathrm{out}}/D_{\beta_\mathrm{s}}    $
does not seem to be decreasing in the case of (d),
where both $ H_\mathrm{I} $ and $ H_\mathrm{Q} $ are integrable.
Especially,
in the case of (a) and (b),
$ D^U_{\mathrm{out}}/D_{\beta_\mathrm{s}}   $ becomes exactly zero even 
at finite $N$, at which eigenstates with $W_j >0$ completely vanish.
This is regarded as a stronger version of the second law at the level of individual energy eigenstates,
which is analogous to the strong ETH.  

Apparently,
$ D^U_{\mathrm{out}}/D_{\beta_\mathrm{s}}    $
does not seem to be decreasing in the case of (d),
where both $ H_\mathrm{I} $ and $ H_\mathrm{Q} $ are integrable.
However, 
as will be shown in inequality \eqref{eq:bound},
which is applicable to this case,
$ D^U_{\mathrm{out}}/D_{\beta_\mathrm{s}}   $ must decrease at least exponentially.
This is because the parameters in inequality \eqref{eq:bound}
in our numerical setup
are very small ($ \beta_{\mathrm{s}}=0.02/J$ and $ \delta=0.001J $),
and therefore the decay rate is too small 
to be observed in our numerical result.
We note that in the case of (d),
$ D^U_{\mathrm{out}}/D_{\beta_\mathrm{s}}   $ itself increases exponentially with $N$,
because $ D_{\beta_\mathrm{s}}  $ itself increases exponentially.
Therefore, $ D^U_{\mathrm{out}} $ does not become exactly zero at finite $N$, which is analogous to the weak ETH in integrable systems
\cite{Biroli2010,Yoshizawa2017,Tasaki2015,Mori2016}.

Table~\ref{table:summary} summarizes
whether the stronger version of the second law
(i.e., $ D^U_{\mathrm{out}} = 0$ at finite $N$)
holds or not in our numerical calculations.
In the case of (c),
whether this property holds is not very clear from Fig.~\ref{fig:scaling} (c),
due to the limitation of our numerical resource.
However,
we argue that
$D^U_{\mathrm{out}}/D_{\beta_\mathrm{s}}   =0$ holds
at finite $N$ in this case too,
from the following discussion and the numerical result
in Fig.~\ref{fig:ETH} (c).

\begin{table}
	\centering
	\caption{\label{table:summary}
		The attainability of $D^U_{\mathrm{out}}/D_{\beta_\mathrm{s}}  =0  $ at finite $N$
		for the four cases of the integrability.
		In the case of (c),
		Fig.~\ref{fig:scaling} does not show
		$D^U_{\mathrm{out}}/D_{\beta_\mathrm{s}}  =0  $,
		while it is expected to be true from the ETH argument
		and the numerical result in Fig.~\ref{fig:ETH}~(c).
	}
	\begin{tabular}{|l|c|c|c|}
		\hline
		\multicolumn{2}{|c|}{}& \multicolumn{2}{c|}{Pre-quench Hamiltonian $ H_\mathrm{I} $} \\ 
		\cline{3-4}
		\multicolumn{2}{|c|}{} & Non-integrable & Integrable \\ 
		\cline{1-4}
		Post-quench & Non-integrable	& (a) True& (b) True  \\ 
		\cline{2-4}
		Hamiltonian $ H_\mathrm{Q} $& Integrable	&  (c) True & (d) False \\ 
		\hline 
	\end{tabular} 
\end{table}

\section{Eigenstate second law from the strong ETH}
\label{sec:ETH}

We discuss
the physical mechanism  of
$D^U_{\mathrm{out}}/D_{\beta_\mathrm{s}}  =0$ on the basis of the strong ETH.

\subsection{Eigenstate second law from the strong ETH}
We assume that
Hamiltonian $ H_{\mathrm{I}} $ satisfies the strong ETH
for observable $ H_{\mathrm{I}}(\tau):=e^{iH_\mathrm{Q}\tau}H_{\mathrm{I}}e^{-iH_\mathrm{Q}\tau} $,
i.e.,
\begin{align}
\braket{j|H_{\mathrm{I}}(\tau)|j}
\simeq
\braket{H_{\mathrm{I}}(\tau)}^{\mathrm{MC}}_{E_j}
\label{eq:ETH}.
\end{align}
Here,
$ \braket{\cdots}^{\mathrm{MC}}_{E_j} $
represents the microcanonical average with respect to $ H_{\mathrm{I}} $
around the energy $ E_j $.

Under this assumption, 
the average work
extracted from $ \ket{j} $ is given by
\begin{align}
 W_j\simeq E_j-\braket{H_{\mathrm{I}}(\tau)}^{\mathrm{MC}}_{E_j}.
\end{align}
Meanwhile,
the average work
from the Gibbs state of $ H_{\mathrm{I}} $ at the mean energy  $ E_j $ is given by
\begin{align}
W_{\mathrm{Gibbs}}:=E_j-\braket{H_{\mathrm{I}}(\tau)}^{\mathrm{Gibbs}}_{E_j}
\simeq 
E_j-\braket{H_{\mathrm{I}}(\tau)}^{\mathrm{MC}}_{E_j}
,
\end{align}
where we used the strong ETH
and the concentration of the energy in the Gibbs state \cite{Tasaki2018}.
We thus obtain $ W_j\simeq W_{\mathrm{Gibbs}} $,
and the passivity of the Gibbs state \cite{Pusz1978,Lenard1978}
(i.e., $ W_{\mathrm{Gibbs}} \leq 0 $)
immediately leads to $  W_j\lesssim0 $.
Therefore,
we conclude that $  W_j\lesssim0 $ for all $ j $
in the case of (a), (b), and (c),
where  the strong ETH holds.

\subsection{Strong ETH for a non-local observable}
We numerically confirm
the strong ETH~\eqref{eq:ETH}.
We remark that in general $ H_{\mathrm{I}}(\tau) $ is a non-local observable,
which involves products of $\mathcal{O}(L)$ local operators. 
The strong ETH for such a non-local many-body observable has not been fully investigated so far.

We calculate
the ETH indicator
$ r_j:=|\braket{j+1| H_{\mathrm{I}}(\tau)|j+1}-\braket{j| H_{\mathrm{I}}(\tau)|j}|/N $,
which was introduced
in Ref.~\cite{Kim2014}.
Here,
$ \ket{j} $ is the $ j $th energy eigenstate
in the increasing order of $E_j$ in the Hilbert space without dividing it into momentum sectors.
If the strong ETH holds,
all of $ r_j $ decay exponentially with $ N $.
In fact,
if there is an athermal eigenstate $\ket{j}$,
the corresponding $ r_j $ does not decay.
However,
even in such a case,
the average of $ r_j $ decays
if the weak ETH holds.

Since we are interested in
eigenstates at positive temperature,
we remove eigenstates
with $ E_j/N>u_\mathrm{Gibbs}(\beta=0) $.
We also remove eigenstates at the edge of the spectrum,
because the density of states is too small there.
Therefore,
we calculated
$ r_j $
for eigenstates satisfying $ E_0/2N\leq E_j/N\leq  u_\mathrm{Gibbs}(\beta=0) $.

\begin{figure}
	\centering
	\includegraphics[width=1.0\linewidth]{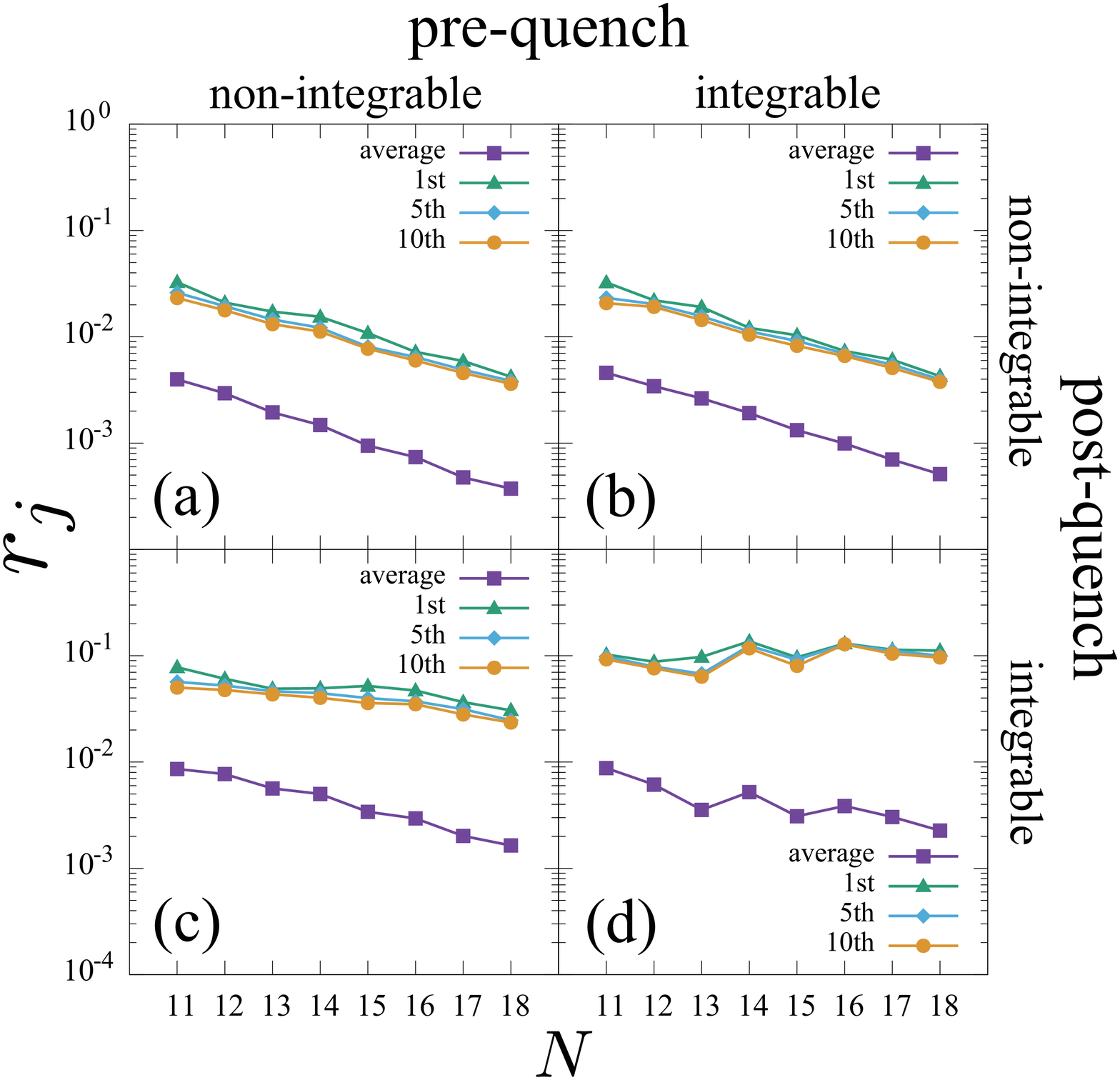}
	\caption{\label{fig:ETH}
		Numerical validation of the ETH of the non-local observable $H_{\mathrm{I}}(\tau) $ with respect to the Hamiltonian $ H_{\mathrm{I}}$. 
		The system-size dependences of
		the average, the largest, the fifth largest, and the tenth largest values of $ r_j $
		are shown for the four cases of the integrability.
		The parameters of Hamiltonians are the same as 
		those used in Fig.~\ref{fig:quench_Ising}.
		The weak ETH is true for all the cases,
		while the strong ETH is true only for (a), (b), and (c).
	}
\end{figure}

Figure \ref{fig:ETH} shows the system-size dependence of
the average, the largest, the fifth largest, and the tenth largest values of $ r_j $.
The average of $ r_j $
decays exponentially in all the cases,
which implies the weak ETH.
On the other hand,
other values of $ r_j $ decay only
in the case of (a), (b), and (c),
where one or both of Hamiltonians $  H_{\mathrm{I}} $ and $  H_{\mathrm{Q}} $ are non-integrable.  
These values do not decay
in the case of (d)
where  both $  H_{\mathrm{I}} $ and $  H_{\mathrm{Q}} $ are integrable.
Therefore,
the strong ETH is valid for the case of (a), (b), and (c),
which is consistent with Table~\ref{table:summary}.

\section{General bound on $ D^U_{\mathrm{out}}/D_{\beta_\mathrm{s}}  $}
\label{sec:bound}

We make an analytical argument about a general bound on  $ D^U_{\mathrm{out}}/D_{\beta_\mathrm{s}}  $.

\subsection{Assumptions and the statement}

We consider a quantum many-body system of $ N $ spins,
and 
denote the number of energy eigenstates of $ H_{\mathrm{I}} $
satisfying  $ E_j\leq E $ by $ D(E) $.
We then assume that
there exists an $ N $-independent positive function $ \sigma(u) $
such that
\begin{align}
\log D(uN) = N\sigma (u) 
+ o(N) 
\label{eq:assumption}
\end{align}
for any $ u\in [u_\mathrm{s}-\delta,u_\mathrm{s}]  $,
where
$ u_\mathrm{s}:= u_\mathrm{Gibbs}(\beta_{\mathrm{s}})$, and
$ o(N) $ satisfies $ o(N)/N\to0 $ in $ N\to\infty $.
From the Boltzmann formula \cite{Ruelle1999},
$\sigma (u)$ is nothing but the entropy density.
We also assume that
$ \sigma'(u_\mathrm{s})=\beta_{\mathrm{s}}>0 $ and $\sigma''(u_\mathrm{s})<0$.
These assumptions are indeed provable in rigorous statistical mechanics
if the system is translation invariant and the interaction is local
\cite{Ruelle1999,Tasaki2018}.

In this setup,
we can show that for an arbitrary fixed unitary operator $ U $
and arbitrary $ \delta>0 $,
the inequality
\begin{align}
\frac{D^U_{\mathrm{out}}}{D_{\beta_\mathrm{s}}}
\lesssim e^{-N\beta_{\mathrm{s}}\delta/2+o(N)}
	\label{eq:bound}
\end{align}
holds.
We show a derivation in the next subsection.
Inequality~\eqref{eq:bound} implies that
the ratio of energy eigenstates,
from which one can extract macroscopic work,
decays at least exponentially with the system size. 
This bound holds for both integrable and non-integrable systems,
as long as the aforementioned assumptions are satisfied.

We also remark a similarity between the above argument and the weak ETH,
which states that
the ratio of athermal eigenstates decays at least exponentially with the system size
\cite{Yoshizawa2017,Tasaki2015,Mori2016}.
Despite this similarity,
the logic behind inequality~\eqref{eq:bound} is different from that behind the weak ETH,
as seen in the derivation of inequality~\eqref{eq:bound}.

\subsection{Derivation of a general bound on $ D^U_{\mathrm{out}}/D_{\beta_\mathrm{s}} $}
\label{sec:proof}
We now derive inequality~\eqref{eq:bound}.
We first consider a slightly different quantity from $ D^U_{\mathrm{out}} $,
denoted by $ \tilde{D}^U_{\mathrm{out}} $,
and prove a rigorous bound on $ \tilde{D}^U_{\mathrm{out}} $.
Then,  inequality~\eqref{eq:bound} is approximately derived as a corollary.

First of all,
we define the stochastic work
\cite{Esposito2009,Campisi2011,Sagawa2012}.
We perform projective measurements of $ H_{\mathrm{I}} $
before and after the operation $ U $,
which give outcomes $ E_j $ and $ E_k$, respectively.
The stochastic work is defined as the difference between the measured energies:
$ w_{j,k}:=E_j-E_k $.
If the initial density operator is $ \rho $,
the joint probability of observing such outcomes is
$ p(j,k):=|\braket{k|U|j}|^2\braket{j|\rho|j} $.

For a given positive constant $ \delta $,
the probability that we extract work larger than $ N\delta $ is given by
$ P(w\geq N\delta):=\sum_{j,k}\theta(w_{j,k}-N\delta)p(j,k) $,
where $ \theta(\cdot) $ is the step function,
i.e., $ \theta(x)=0\ (x\leq 0)$ and $\ \theta(x)=1\ (x> 0)$.
Here, $ P(w\geq N\delta) $ is
the probability of extracting macroscopic work,
as $N \delta $ is proportional to the system size.
For an energy eigenstate,
the probability of extracting macroscopic work is
$ P_j(w\geq N\delta)=\sum_k\theta(E_j-E_k-N\delta)|\braket{k|U|j}|^2 $.
If  $  P_j(w\geq N\delta) $ is sufficiently small for a small $\delta$, 
we cannot extract macroscopic work 
from the energy eigenstate $ \ket{j} $.

We consider the number of energy eigenstates in $ M_{\beta_\mathrm{s}} $,
from which we can extract macroscopic work 
by a fixed operation $ U $.
We focus on the energy eigenstate $ \ket{j} $ with which
$ P_j(w\geq N\delta)  $ decays exponentially with $ N $.
In other words,
its energy distribution after the protocol shows the large deviation behavior \cite{Touchette2009},
and the probability of observing lower energy than the initial state
is negligibly small on a macroscopic scale.
Then,
we denote by $ \tilde{D}^U_{\mathrm{out}} $
the number of energy eigenstates
whose  $ P_j(w\geq N\delta)  $ is greater than
$ e^{-N\beta_{\mathrm{s}}\delta/2} $:
\begin{align}
\tilde{D}^U_{\mathrm{out}}
:=
\#
\{j\in M_{\beta_\mathrm{s}} : 
P_j(w\geq N\delta)\geq e^{-N\beta_{\mathrm{s}}\delta/2}\}
.
\end{align}
We note that $ e^{-N\beta_\mathrm{s}\delta/2} $ is negligibly small for large $ N $,
which corresponds to the second law as work extraction.

In this setup, we can rigorously prove the following inequality:
\begin{align}
\frac{\tilde{D}^U_{\mathrm{out}}}{D_{\beta_\mathrm{s}}}
\leq
e^{-N\beta_\mathrm{s}\delta/2+o(N)}.
\label{eq:modified_bound}
\end{align}

The proof of this inequality is as follows.
From the definition of 
$ \tilde{D}^U_{\mathrm{out}} $,
we have
\begin{align}
\tilde{D}^U_{\mathrm{out}}
\leq
e^{N\beta_\mathrm{s}\delta/2}
\sum_{j\in M_{\beta_{\mathrm{s}}}}
P_j(w\geq N\delta)
\label{eq:bound_Dout}
.
\end{align}
We can evaluate the sum
in the right-hand side
by using the definition of $ P_j(w\geq N\delta) $
as
\begin{align}
&\sum_{j\in M_{\beta_{\mathrm{s}}}}
P_j(w\geq N\delta)
\nonumber\\
&=
\sum_{j\in M_{\beta_{\mathrm{s}}}}
\sum_k\theta[E_j-E_k-N\delta]
|\braket{E_k|U|E_j}|^2 
\\
&\leq
\sum_{j\in M_{\beta_{\mathrm{s}}}}
\sum_k\theta[( u_\mathrm{s}-\delta)N-E_k]
|\braket{E_k|U|E_j}|^2 
\\
&\leq 
\sum_k\theta[( u_\mathrm{s}-\delta)N-E_k]
\\
&=D(( u_\mathrm{s}-\delta)N),
\label{eq:bound_sum_Pj}
\end{align}
where we used 
$\sum_{j\in M_{\beta_{\mathrm{s}}}}
|\braket{E_k|U|E_j}|^2\leq 1  $
to obtain the fourth line.
From the assumption \eqref{eq:assumption},
we obtain
\begin{align}
\frac{D((u_\mathrm{s}-\delta)N)}{D(u_\mathrm{s}N)}
\leq e^{-N[\sigma(u_\mathrm{s})-\sigma(u_\mathrm{s}-\delta )]+o(N)}. 
\end{align}
Meanwhile,
$ \sigma'(u_\mathrm{s})=\beta_\mathrm{s}$ and $\sigma''(u_\mathrm{s})<0 $
imply
$ \sigma(u_\mathrm{s})>\sigma(u_\mathrm{s}-\delta )+\beta \delta$.
Therefore,
we have
\begin{align}
\frac{D((u_\mathrm{s}-\delta)N)}{D(u_\mathrm{s}N)}
\leq e^{-N\beta_\mathrm{s}\delta+o(N)}. 
\end{align}
By combining this inequality and inequality \eqref{eq:bound_sum_Pj},
we have
\begin{align}
\frac{\tilde{D}^U_{\mathrm{out}}}{D_{\beta_\mathrm{s}}}
\leq
\frac{e^{N\beta_\mathrm{s}\delta/2}D((u_\mathrm{s}-\delta)N)}{D(u_\mathrm{s}N)}
\leq e^{-N\beta_\mathrm{s}\delta/2+o(N)},
\end{align}
which implies inequality \eqref{eq:modified_bound}.

The definition of $ \tilde{D}^U_{\mathrm{out}} $
is slightly different from
$ D^U_{\mathrm{out}} $. 
For the latter,
we can only obtain an approximate inequality \eqref{eq:bound} as follows.
If $ P_j(w\geq N\delta)< e^{-N\beta_\mathrm{s}\delta/2} $,
the average work extraction from $ \ket{j} $ is bounded as
\begin{align}
W_j
&= \int_{w< N\delta}wP_j(w)dw
+\int_{w\geq N\delta}wP_j(w)dw
\\
&= \int_{w< N\delta}wP_j(w)dw + e^{-\mathcal{O}(N)}
\\
&\lesssim N\delta.
\end{align}
From this inequality,
we have
$ \tilde{D}^U_{\mathrm{out}}\gtrsim D^U_{\mathrm{out}} $.
Therefore,
we obtain
\begin{align}
\frac{D^U_{\mathrm{out}} }{D_{\beta_\mathrm{s}}}
\lesssim 
\frac{ \tilde{D}^U_{\mathrm{out}}}{D_{\beta_\mathrm{s}}}
\leq e^{-N\beta_\mathrm{s}\delta/2+o(N)}.
\end{align}

\subsection{Work extraction from a general pure state}
\label{sec:general_state}
From inequality~\eqref{eq:bound},
we can show that a positive amount of work cannot be extracted from any pure state
if it has sufficiently large effective dimensions.
We consider a general pure state
in the Hilbert space of positive temperature region,
i.e.,
$ \ket{\psi}=\sum_{j\in M_{\beta_{\mathrm{s}}}} c_j\ket{j}$ with $  \sum_{j\in M_{\beta_{\mathrm{s}}}}|c_j|^2=1 $.
We apply
the general work extraction protocol in Sec.~\ref{sec:proof}
to $ \ket{\psi} $.

The probability of extracting macroscopic work
from $ \ket{\psi} $ is 
$ P(w\geq N\delta)=\sum_{j\in M_{\beta_{\mathrm{s}}}} |c_j|^2 P_j(w\geq N\delta)$.
The Cauchy-Schwartz inequality
gives
\begin{align}
&P(w\geq N\delta)
\nonumber\\
&\leq
\sqrt{\left(\sum_{j\in M_{\beta_{\mathrm{s}}}}{P_j(w\geq N\delta)}^2\right)
	\left(\sum_{j\in M_{\beta_{\mathrm{s}}}}|c_j|^4\right)}
\\
&
\leq
\sqrt{
	\frac{
	D_{\beta_{\mathrm{s}}} e^{-N\beta_{\mathrm{s}}\delta/2+o(N)}
	+
	D_{\beta_{\mathrm{s}}} (1-e^{-N\beta_{\mathrm{s}}\delta/2+o(N)})
	e^{-N\beta_{\mathrm{s}}\delta}
}
{
		D_{\mathrm{eff}}}
}
\\
&\leq
\sqrt{\frac{D_{\beta_{\mathrm{s}}} e^{-N\beta_{\mathrm{s}}\delta/2+o(N)}}{D_{\mathrm{eff}}}},
\end{align}
where we used
inequality~\eqref{eq:modified_bound} to obtain the third line
and
defined the effective dimension
$D_{\mathrm{eff}}
:=
1/\sum_{j\in M_{\beta_{\mathrm{s}}}}|c_j|^4 $~\cite{Reimann2008}.
From this inequality,
the average work extraction from $ \ket{\psi} $
is bounded as
\begin{align}
W
&= \int_{w< N\delta}wP(w)dw
+\int_{w\geq N\delta}wP(w)dw
\\
&\lesssim  N\delta + w_{\mathrm{max}}\sqrt{\frac{D_{\beta_{\mathrm{s}}} e^{-N\beta_{\mathrm{s}}\delta/2+o(N)}}{D_{\mathrm{eff}}}},
\end{align}
where $ w_{\mathrm{max}} $ is the maximum value of the stochastic work.
$ w_{\mathrm{max}} $ scales at most linearly with $ N $.

Therefore,
$ W \lesssim  N\delta$
holds for any unitary operation,
if the initial state has 
sufficiently large effective dimensions:
$ D_{\mathrm{eff}}\gg  D_{\beta_{\mathrm{s}}}e^{-N\beta_{\mathrm{s}}\delta/2}$.
Because the effective dimension of
a typical pure state with respect to the Haar measure is estimated as
$ D_{\mathrm{eff}}\simeq D_{\beta_{\mathrm{s}}} \gg  D_{\beta_{\mathrm{s}}} e^{-N\beta_{\mathrm{s}}\delta/2}$~\cite{Linden2009},
we cannot extract macroscopic work
from a typical pure state.

\section{Conclusion}
\label{sec:conclusion}
In this paper, we have investigated work extraction
from a single energy eigenstate by a specific cyclic operation
and clarified the effect of the integrability on work extraction.
Our main numerical results are presented in Fig.~\ref{fig:scaling} and Table~\ref{table:summary}.

Roughly speaking,
work extraction by our quench protocol
is impossible from \textit{any} energy eigenstate
if the system is non-integrable,
while it is impossible from \textit{most} energy eigenstates
if the system is integrable.
We thus conjecture that
the same result would hold for a much broader class of 
physical work extraction protocols
with local and translation invariant Hamiltonians.
This is in fact feasible in light of the fact that
the strong ETH has been confirmed to be true for a broad class of Hamiltonians
\cite{Jensen1985,Rigol2008,Steinigeweg2013,Steinigeweg2014,Beugeling2014,Kim2014,Khodja2015,Garrison2015,Mondaini2016,Dymarsky2018,Yoshizawa2017}.
If the above conjecture is true,
we can conclude that the second law is so universal
that it applies at the level of individual energy eigenstates.
Further investigation of this conjecture is a future issue.

On the other hand,
in many-body localization systems (MBL) without translation invariance
\cite{Altman2014,Nandkishore2015,Abanin2018},
not only the strong ETH but also the weak ETH fails \cite{Pal2010,Imbrie2016}.
The possibility of  work extraction from such an exotic system
is an interesting direction for future investigation.

\begin{acknowledgments}
K.K. is grateful to K.~Yamaga for valuable discussions.
K.K.  are supported by JSPS KAKENHI Grant Number JP17J06875.
E.I. and T.S. are supported by JSPS KAKENHI Grant Number JP16H02211.
E.I. is also supported by JSPS KAKENHI Grant Number JP15K20944.
\end{acknowledgments}


\appendix
\section{Level statistics}
\label{sec:level_statistics}
In order to confirm
that the parameters chosen in the numerical calculation have sufficient non-integrability,
we investigate the level spacing statistics.
In particular,
we calculated the ratio of adjacent energy gap~\cite{Oganesyan2007}
which is defined as
$ r_j:=\min(\Delta_{j+1},\Delta_{j})/\max(\Delta_{j+1},\Delta_{j})$	.
Here,
$  \Delta_j:=E_{j+1}-E_j  $
is the gap between the eigenenergies,
and
the eigenenergies are listed in increasing order
in each momenta sector.
If the system is nonintegrable,
$ r_j $ follows the Wigner-Dyson distribution
given by random matrix theory.
For the system with time-reversal symmetry,
the distribution is well fitted by the Gaussian orthogonal ensemble (GOE),
whose explicit form in our formulation is given by~\cite{Atas2013}:
\begin{align}
	P_{\mathrm{GOE}}(r)=\frac{27}{4}\frac{r+r^2}{(1+r+r^2)^{\frac{5}{2}}}\Theta(1-r).
\end{align}
By numerical exact diagonalization,
we obtained the distributions
of $ r_j $ for all the momentum sectors,
except for $ k=0$ and  $k=\pi $.
We denote the average distribution of $ r_j $ over 
these momentum sectors by $ P(r) $.
We consider an indicator of the distance between
$ P(r) $ and $ P_{\mathrm{GOE}}(r) $~\cite{Santos2010}:
\begin{align}
	\alpha:=
	\frac{\sum_n|P(n\Delta r)-P_{\mathrm{GOE}}(n\Delta r)|}{\sum_nP_{\mathrm{GOE}}(n\Delta r)},
	\label{seq:GOE}
\end{align}
where $ \Delta r $ is the size of a bin used in our calculation of  $ P(r) $.
This indicator is expected to be small if the system is sufficiently non-integrable.

\begin{figure}
	\centering
	\includegraphics[width=0.8\linewidth]{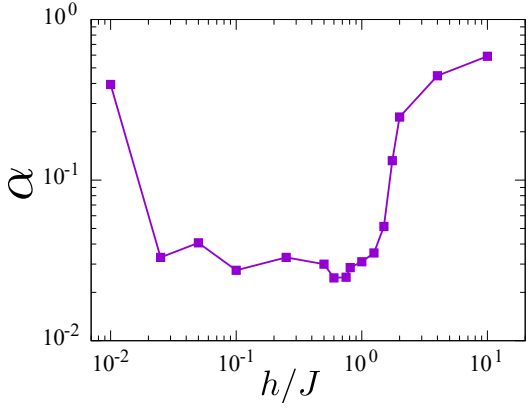}
	\caption{\label{fig_level_statistics}
		The $ h $-dependence of  the distance $ \alpha $
		between the distribution of the ratio of consecutive energy gaps
		and that of the GOE prediction [Eq.~\eqref{seq:GOE}].
		The other parameters are fixed to $ J=1, J'=0, g=(\sqrt{5}+5)/8, N=18$.
		Our calculation was performed
		using the central half of the spectrum,
		where the width of the bin is $ \Delta r=10^{-2} $.
	}
\end{figure}

Figure~\ref{fig_level_statistics} shows
$ \alpha $ for various $ h $
with fixing the other parameters to $ J=1, J'=0, g=\frac{\sqrt{5}+5}{8}, N=18 $.
From Fig.~\ref{fig_level_statistics},
we find
that the difference between $ P(r) $ and $ P_{\mathrm{GOE}} $
is small enough for $ 2.5\times 10^{-2}\lesssim h\lesssim 1.5$.
Therefore,
we conclude
that our model is sufficiently non-integrable
for $ 2.5\times 10^{-2}\lesssim h\lesssim 1.5$.

\section{The time dependence of the average work}
\label{sec:time_dependence}

\begin{figure}
	\centering
	\includegraphics[width=1.0\linewidth]{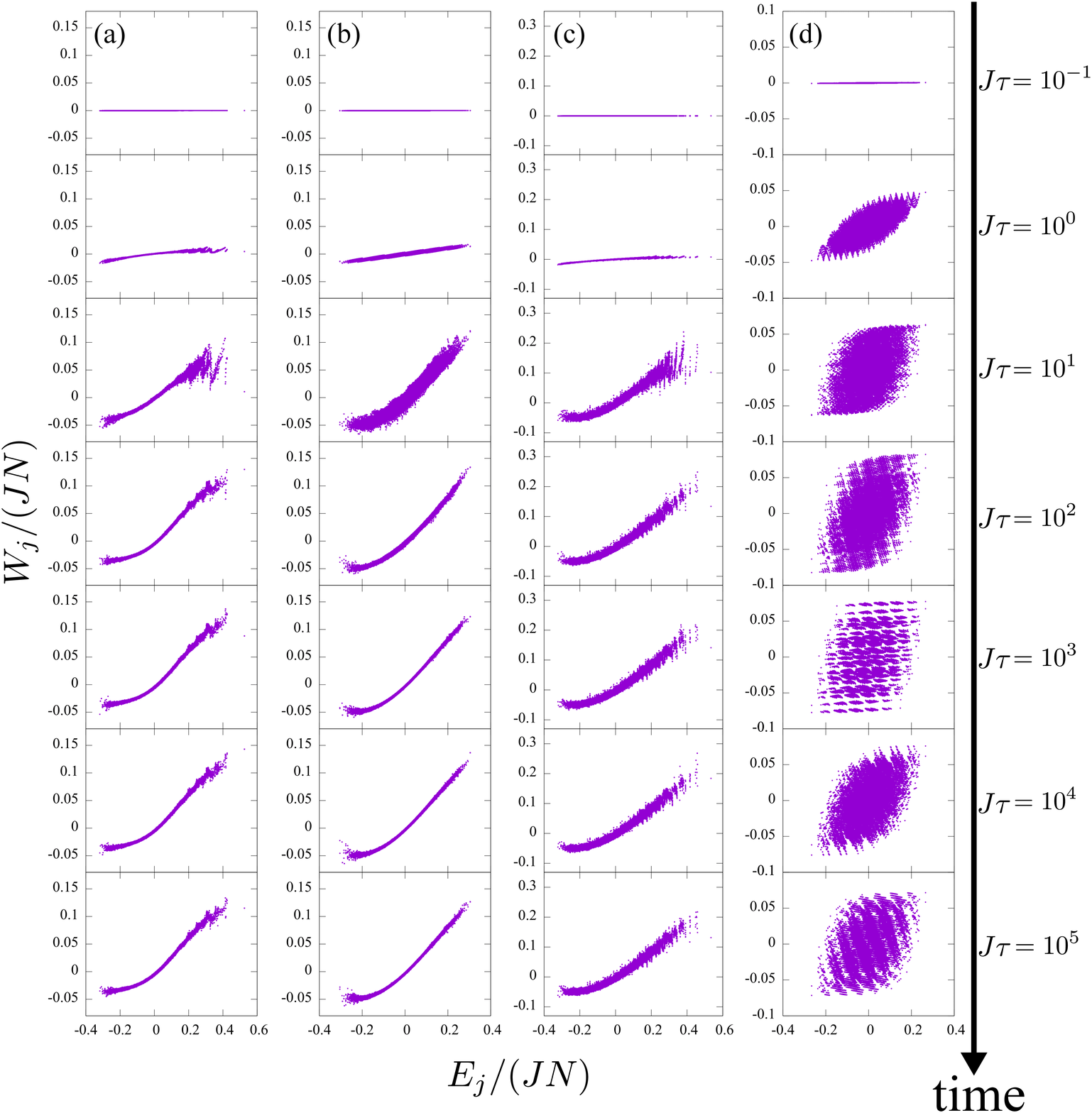}
	\caption{\label{fig_work_time}
		The $ \tau $-dependence of $ W_j/N $ with $ N=18 $
		and $ \tau=10^{-1}, 10^{0}, 10^{1}, 10^{2}, 10^{3},10^{4}, 10^{5}$.
		Quench from
		(a) non-integrable to non-integrable
		($ J=J'=1, g_{\mathrm{I}}=g_{\mathrm{Q}}=(\sqrt{5}+5)/8, h_{\mathrm{I}}=0\to h_{\mathrm{Q}}= 1.0 $),
		(b) integrable to non-integrable
		($ J=1, J'=0, g_{\mathrm{I}}=g_{\mathrm{Q}}=(\sqrt{5}+5)/8, h_{\mathrm{I}}=0\to h_{\mathrm{Q}}=1.0$),
		(c) non-integrable to integrable
		($ J=1, J'=0, g_{\mathrm{I}}=g_{\mathrm{Q}}=(\sqrt{5}+5)/8, h_{\mathrm{I}}=1.0 \to h_{\mathrm{Q}}=0 $),
		and
		(d) integrable to integrable
		($ J=1, J'=0, h=0, g_{\mathrm{I}}=0.5\to g_{\mathrm{Q}}=1.5 $).
	}
\end{figure}

We consider the dependence of the average work $W_j$ on the waiting time $\tau$
in the protocol of Fig.~\ref{fig:quench_Ising}.
In the following, (a), (b), (c), (d) indicate the four cases of the integrability,
as defined in Sec.~\ref{sec:numerical} .
Figure~\ref{fig_work_time} shows the average work per qubit
for the full spectrum for various $\tau$.
For $ \tau=10^{-1} $,
$ W_j/N $ is almost zero for all the cases
and takes a non-zero value
after $ \tau=10^{-1} $.
$ W_j/N $
takes an almost constant value up to temporal fluctuations
after $ \tau=10^2 $.
The temporal fluctuations are small for the three cases
(a), (b), and (c),
and
the appearances of the scatter plots of $ W/N $
are almost the same for $ \tau=10^2,10^3, 10^4,10^5 $.
On the other hand,
the temporal fluctuations are  larger in the case of  (d), 
and the appearances of the scatter plots are not the same
for $ \tau=10^2,10^3, 10^4,10^5 $.

\section{The parameter-dependence of $ D^U_{\mathrm{out}}/D_{\beta_\mathrm{s}} $}
\label{sec:parameter_dependence}

In this Appendix,
we show the parameter-dependence of 
$D^U_{\mathrm{out}}/D_{\beta_\mathrm{s}} $.
We change only one parameter at once
and fix the other parameters
to the same value as in Fig.~\ref{fig:scaling}.
The parameters used in Fig.~\ref{fig:scaling} are given by
$ J\tau=10^4, J\beta_{\mathrm{s}}=0.02$, and  $\delta/J=0.001$.

\subsection{The $ \beta_{\mathrm{s}} $-dependence}

\begin{figure}
	\includegraphics[width=1.0\linewidth]{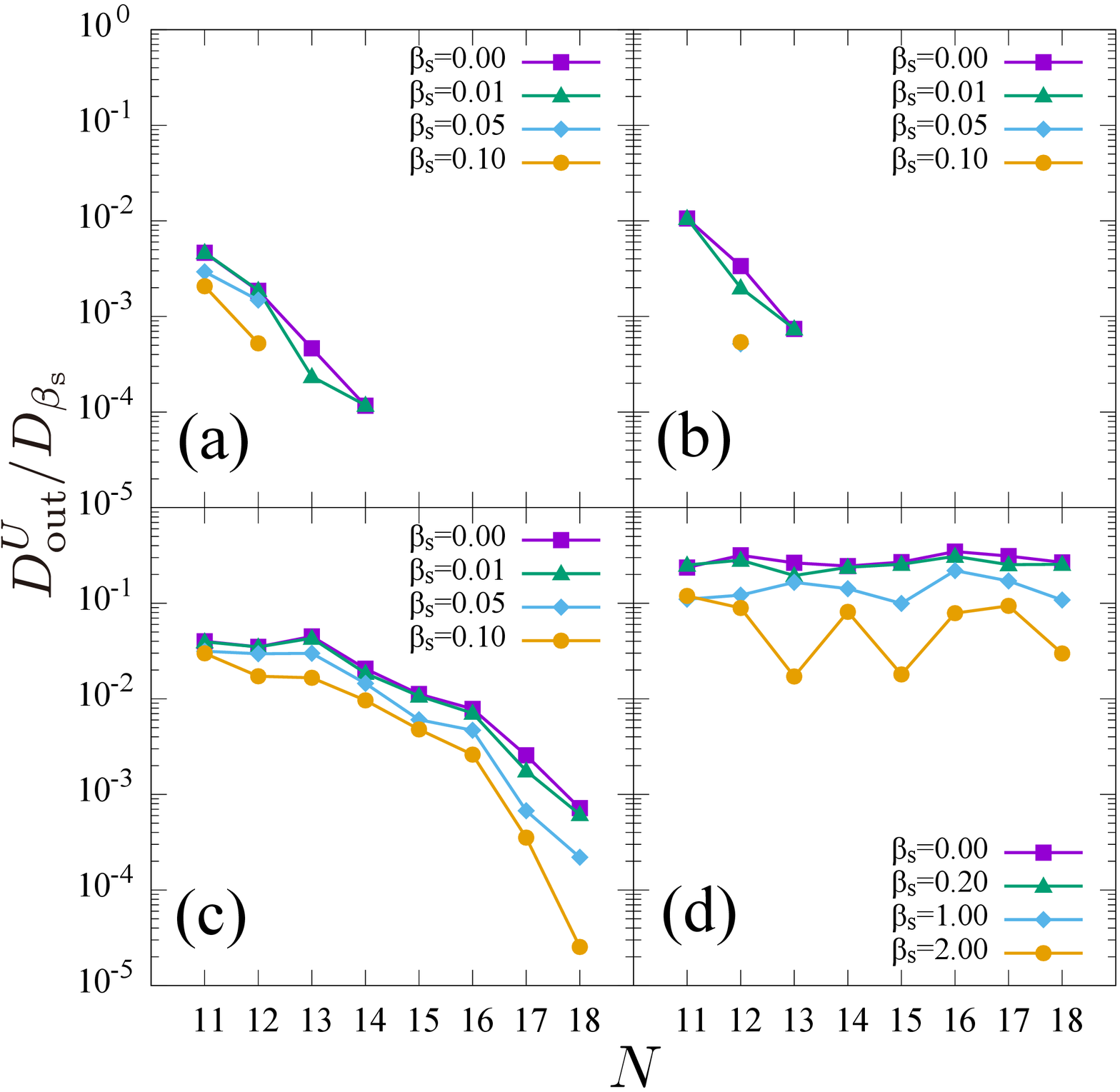}
	\caption{\label{fig:betas_dependence}
		The $ N $-dependence of $ D^U_{\mathrm{out}}/D_{\beta_\mathrm{s}}    $
		for various $ \beta_{\mathrm{s}} $.
		Quench from
		(a) non-integrable to non-integrable
		($ J=J'=1, g_{\mathrm{I}}=g_{\mathrm{Q}}=(\sqrt{5}+5)/8, h_{\mathrm{I}}=0\to h_{\mathrm{Q}}= 1.0 $),
		(b) integrable to non-integrable
		($ J=1, J'=0, g_{\mathrm{I}}=g_{\mathrm{Q}}=(\sqrt{5}+5)/8, h_{\mathrm{I}}=0\to h_{\mathrm{Q}}=1.0$),
		(c) non-integrable to integrable
		($ J=1, J'=0, g_{\mathrm{I}}=g_{\mathrm{Q}}=(\sqrt{5}+5)/8, h_{\mathrm{I}}=1.0 \to h_{\mathrm{Q}}=0 $),
		and
		(d) integrable to integrable
		($ J=1, J'=0, h=0, g_{\mathrm{I}}=0.5\to g_{\mathrm{Q}}=1.5 $).
	}
\end{figure}

Figure~\ref{fig:betas_dependence} shows
the $ \beta_{\mathrm{s}} $-dependence of $ D^U_{\mathrm{out}}/D_{\beta_\mathrm{s}}   $.
For the two cases (a) and (b),
where $ H_\mathrm{Q} $ is non-integrable,
$ D^U_{\mathrm{out}}/D_{\beta_\mathrm{s}}    $ vanishes
at finite $ N $ regardless of $ \beta_{\mathrm{s}} $.
In the case of  (c),
$ D^U_{\mathrm{out}}/D_{\beta_\mathrm{s}}    $
with  large $ \beta_{\mathrm{s}} $
decays faster,
whereas in the case of  (d),
we do not observe the decay of $ D^U_{\mathrm{out}}/D_{\beta_\mathrm{s}}    $.

\subsection{The $ \delta $-dependence}

\begin{figure}
	\centering
	\includegraphics[width=1.0\linewidth]{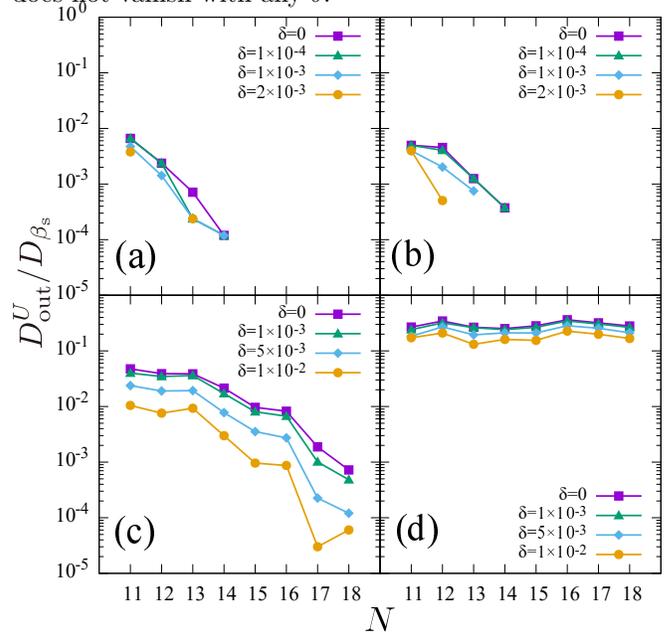}
	\caption{\label{fig:delta_dependence}
		The $ N $-dependence of $ D^U_{\mathrm{out}}/D_{\beta_\mathrm{s}}    $
		for various $ \delta $.
		The parameters of the Hamiltonians are
		the same as those used in Fig.~\ref{fig:betas_dependence}.
	}
\end{figure}

Figure~\ref{fig:delta_dependence} shows
the  $ \delta $-dependence of $ D^U_{\mathrm{out}}/D_{\beta_\mathrm{s}}    $.
For the two cases (a) and (b),
$ D^U_{\mathrm{out}}/D_{\beta_\mathrm{s}}    $ vanishes
at finite $ N $ regardless of $ \delta $.
In the case of (c),
$ D^U_{\mathrm{out}}/D_{\beta_\mathrm{s}}    $
with  large $ \delta $
decays faster,
and 
$ D^U_{\mathrm{out}}/D_{\beta_\mathrm{s}}    $ also vanishes
at finite $ N $ when $ \delta=0.02 $.
However,
in the case of (d),
$ D^U_{\mathrm{out}}/D_{\beta_\mathrm{s}}    $ does not vanish with any $ \delta $.

\newpage

\subsection{Summary}
From the above results,
we conclude that our statement in the main text is independent of the choice of the parameters:
$ D^U_{\mathrm{out}}/D_{\beta_\mathrm{s}}   $ vanishes
at finite $ N $ for (a) and (b)
and
does not vanishes for (d).
The case of (c) is marginal, 
which is again consistent with our argument in the main text.



\begin{thebibliography}{99}
	
\bibitem{Tasaki2000stat}
H. Tasaki, arXiv:cond-mat/0009206 (2000).

\bibitem{Allahverdyan2002}
A.~E.~Allahverdyan and T.~M.~Nieuwenhuizen, Phys.~A \textbf{305}, 542 (2002).
	

\bibitem{Pusz1978}
W.~Pusz and S.~L.~Woronowicz, Comm.~Math.~Phys. \textbf{58}, 273 (1978).

\bibitem{Lenard1978}
A.~Lenard, J.~Stat.~Phys. \textbf{19}, 575 (1978).


\bibitem{Polkovnikov2011}
A.~Polkovnikov, K.~Sengupta, A.~Silva, and M.~Vengalattore, Rev.~Mod.~Phys. \textbf{83}, 863 (2011).

\bibitem{DAlessio2015}
L. D'Alessio, Y. Kafri, A. Polkovnikov, and M. Rigol, Adv. Phys. \textbf{65}, 239 (2016).

\bibitem{Eisert2015}
J. Eisert, M. Friesdorf, and C. Gogolin, Nat. Phys. \textbf{11}, 124 (2015).

\bibitem{Gogolin2016}
C. Gogolin and J. Eisert, Rep. Prog. Phys. \textbf{79}, 056001 (2016).


\bibitem{Berry1977}
M. V. Berry, J. Phys. A \textbf{10}, 2083 (1977).

\bibitem{Peres1984}
A. Peres, Phys. Rev. A \textbf{30}, 504 (1984).


\bibitem{Deutsch1991}
J. M. Deutsch, Phys. Rev. A \textbf{43}, 2046 (1991).

\bibitem{Srednicki1994}
M. Srednicki, Phys. Rev. E \textbf{50}, 888 (1994).

\bibitem{Rigol2008}
M. Rigol, V. Dunjko, and M. Olshanii, Nature (London) \textbf{452}, 854 (2008).


\bibitem{Jensen1985}
R. V. Jensen and R. Shankar, Phys. Rev. Lett. \textbf{54}, 1879 (1985).

\bibitem{Steinigeweg2013}
R. Steinigeweg, J. Herbrych, and P. Prelov\^{s}ek, Phys. Rev. E \textbf{87}, 012118 (2013).

\bibitem{Steinigeweg2014}
R. Steinigeweg, A. Khodja, H. Niemeyer, C. Gogolin, and J. Gemmer, Phys. Rev. Lett. \textbf{112}, 130403 (2014).

\bibitem{Beugeling2014}
W. Beugeling, R. Moessner, and M. Haque, Phys. Rev. E \textbf{89}, 042112 (2014).

\bibitem{Kim2014}
H. Kim, T. N. Ikeda, and D. A. Huse, Phys. Rev. E \textbf{90}, 052105 (2014).

\bibitem{Khodja2015}
A. Khodja, R. Steinigeweg, and J. Gemmer, Phys. Rev. E \textbf{91}, 012120 (2015).


\bibitem{Mondaini2016}
R. Mondaini, K. R. Fratus, M. Srednicki, and M. Rigol, Phys. Rev. E \textbf{93}, 032104 (2016).

\bibitem{Garrison2015}
J. R. Garrison and T. Grover, Phys. Rev. X \textbf{8}, 021026 (2018).

\bibitem{Dymarsky2018}
A. Dymarsky, N. Lashkari, and H. Liu, Phys. Rev. E \textbf{97}, 012140 (2018).

\bibitem{Yoshizawa2017}
T. Yoshizawa, E. Iyoda, and T. Sagawa, Phys. Rev. Lett. \textbf{120}, 200604 (2018).



\bibitem{Rigol2009}
M. Rigol, Phys. Rev. Lett. \textbf{103}, 100403 (2009).

\bibitem{Biroli2010}
G. Biroli, C. Kollath, and A. M. L\"{a}uchli, Phys. Rev. Lett. \textbf{105}, 250401 (2010).

\bibitem{Alba2015}
V. Alba, Phys. Rev. B \textbf{91}, 155123 (2015).

\bibitem{Essler2016}
F. H. L. Essler and M. Fagotti, J. Stat. Mech. 064002 (2016).

\bibitem{Vidmar2016}
L. Vidmar and M. Rigol, J. Stat. Mech. 064007 (2016).


\bibitem{Tasaki2015}
H. Tasaki, J. Stat. Phys. \textbf{163}, 937 (2016).

\bibitem{Mori2016}
T. Mori, arXiv:1609.09776 (2016).

\bibitem{Iyoda2016}
E. Iyoda, K. Kaneko, and T. Sagawa, Phys. Rev. Lett. \textbf{119}, 100601 (2017).


\bibitem{Tasaki2000pure}
H. Tasaki, arXiv:cond-mat/0011321 (2000).

\bibitem{Goldstein2013}
S. Goldstein, T. Hara, and H. Tasaki, arXiv:1303.6393 (2013).

\bibitem{Ikeda2013}
T. N. Ikeda, N. Sakumichi, A. Polkovnikov, and M. Ueda, Ann. Phys. (Amsterdam) \textbf{354}, 338 (2015).

\bibitem{Kaneko2017}
K. Kaneko, E. Iyoda, and T. Sagawa, Phys. Rev. E \textbf{96} 062148 (2017).

\bibitem{Dorner2013}
R.~Dorner, S.~R.~Clark, L.~Heaney, R.~Fazio, J.~Goold, and V.~Vedral, Phys.~Rev.~Lett. \textbf{110}, 230601 (2013).

\bibitem{Gallego2014}
R. Gallego, A. Riera, and J. Eisert, New J. Phys. \textbf{16}, 125009 (2014).

\bibitem{Perarnau-Llobet2016}
M.~Perarnau-Llobet, A.~Riera, R.~Gallego, H.~Wilming, and J.~Eisert, New~J.~Phys. \textbf{18}, 123035 (2016).

\bibitem{Modak2017}
R.~Modak and M.~Rigol, Phys.~Rev.~E \textbf{95}, 062145 (2017).

\bibitem{Le2018}
T.~P.~Le, J.~Levinsen, K.~Modi, M.~M.~Parish, and F.~A. Pollock, Phys.~Rev.~A \textbf{97}, 022106 (2018).

\bibitem{Jin2016}
F. Jin, R. Steinigeweg, H. De Raedt, K. Michielsen, M. Campisi, and J. Gemmer, Phys. Rev. E \textbf{94}, 012125 (2016).

\bibitem{Schmidtke2017}
D.~Schmidtke, L.~Knipschild, M.~Campisi, R.~Steinigeweg, and J.~Gemmer, Phys. Rev. E \textbf{98}, 012123 (2018).



\bibitem{Trotzky2011}
S. Trotzky, Y-A. Chen, A. Flesch, I.~P.~McCulloch, U.~Schollw\"{o}ck J.~Eisert, and I. Bloch, Nat. Phys. \textbf{8}, 325 (2012).

\bibitem{Langen2015}
T. Langen, S. Erne, R. Geiger, B. Rauer, T. Schweigler, M. Kuhnert, W. Rohringer, I. E. Mazets, T. Gasenzer, and J. Schmiedmayer, Science \textbf{348}, 207 (2015).

\bibitem{Neill2016}
C. Neill, P. Roushan, M. Fang, Y. Chen, M. Kolodrubetz, Z. Chen, A. Megrant, R. Barends, B. Campbell, B. Chiaro, A. Dunsworth, E. Jeffrey, J. Kelly, J. Mutus, P. J. J. O'Malley, C. Quintana, D. Sank, A. Vainsencher, J. Wenner, T. C. White, A. Polkovnikov, and J. M. Martinis,
Nat. Phys. \textbf{12}, 1037 (2016).


\bibitem{Ruelle1999}
D. Ruelle, \textit{Statistical Mechanics: Rigorous Results} (World Scientific, Singapore, 1999).

\bibitem{Tasaki2018}
H.~Tasaki,
J. Stat. Phys. \textbf{172}, 905 (2018).

\bibitem{Esposito2009}
M.~Esposito, U.~Harbola, and S.~Mukamel, Rev.~Mod.~Phys. \textbf{81}, 1665 (2009).

\bibitem{Campisi2011}
M.~Campisi, P.~H\"{a}nggi, and P.~Talkner, Rev.~Mod.~Phys. \textbf{83}, 771 (2011).

\bibitem{Sagawa2012}
T.~Sagawa, Lect.~Quantum~Comput.~Thermodyn.~Stat.~Phys. \textbf{8}, 127 (2012).

\bibitem{Touchette2009}
H. Touchette, Phys. Rep. \textbf{478}, 1 (2009).


\bibitem{Reimann2008}
P.~Reimann, Phys.~Rev.~Lett. \textbf{101}, 190403 (2008).

\bibitem{Linden2009}
N.~Linden, S.~Popescu, A.~J.~Short, and A.~Winter, Phys.~Rev.~E \textbf{79}, 061103 (2009).

\bibitem{Nandkishore2015}
R. Nandkishore and D. A. Huse, Annu. Rev. Condens. Matter Phys. \textbf{6}, 15 (2015).

\bibitem{Altman2014}
E. Altman and R. Vosk, Annu. Rev. Condens. Matter Phys. \textbf{6}, 383 (2015).

\bibitem{Abanin2018}
D. A. Abanin, E. Altman, I. Bloch, and M. Serbyn, arXiv:1804.11065 (2018).


\bibitem{Pal2010}
A. Pal and D. A. Huse, Phys. Rev. B \textbf{82}, 174411 (2010).

\bibitem{Imbrie2016}
J. Z. Imbrie, J. Stat. Phys. \textbf{163}, 998 (2016).


\bibitem{Oganesyan2007}
V.~Oganesyan and D.~A.~Huse, Phys.~Rev.~B \textbf{75}, 155111 (2007).

\bibitem{Atas2013}
Y. Y. Atas, E. Bogomolny, O. Giraud, and G. Roux,  Phys.~Rev.~Lett. \textbf{110}, 084101 (2013).

\bibitem{Santos2010}
L. F. Santos and M. Rigol, Phys.~Rev.~E \textbf{81}, 036206 (2010).



\end{thebibliography}
\end{document}